\def\Pp{\mathbb{P}}
\def\Ep{\mathbb{E}}
\def\sect#1{\noindent{\bf #1\unskip}}
\def\grsup{\mathop{\overline{\rm gr}}\nolimits}

\def\gr{\mathop{{\rm gr}}\nolimits}
\documentclass{siam-wns-article}
\usepackage{amsmath,amssymb,amsthm}
\newtheorem{theorem}{Theorem}

\newtheorem{example}{Example}
\title{Exponential decay of connectivity and uniqueness in percolation 
  on finite and infinite graphs}

\author{Kathleen E. Hamilton and Leonid P. Pryadko}

\usehyperref 

\begin{document}
\maketitle

\sect{Summary.}
We give an upper bound for the uniqueness transition on an arbitrary
locally finite graph ${\cal G}$ in terms of the limit of the spectral
radii $\rho\left[ H({\cal G}_t)\right]$ of the non-backtracking (Hashimoto)
matrices for an increasing sequence of subgraphs
${\cal G}_t\subset{\cal G}_{t+1}$ which converge to ${\cal G}$.  With the
added assumption of \emph{strong local connectivity} for the
oriented line graph (OLG) of ${\cal G}$, connectivity on any finite subgraph
${\cal G}'\subset{\cal G}$ decays exponentially for $p<(\rho\left[ H({\cal G}^{\prime})\right])^{-1}$. 
  
\sect{Introduction.}  Percolation is widely used in network theory
applications, yet formation of an infinite cluster is not
sufficient to ensure high likelihood that an arbitrary pair of 
selected sites are connected, since the percolation cluster may not be
unique. In this work we give upper bounds on the connectivity in site
percolation on finite and infinite graphs in terms of the
corresponding non-backtracking (Hashimoto) matrices, and related
bounds for the uniqueness transition.

\sect{Definitions.}  For a graph ${\cal G}$ with the vertex set
${\cal V}\equiv {\cal V}({\cal G})$ and edge set ${\cal E}$, we also
consider the set of arcs (directed edges) ${\cal A}({\cal G})$.  
The Hashimoto\cite{Hashimoto-matrix-1989} matrix $H\equiv H({\cal G})$ is the
adjacency matrix of the oriented line graph of ${\cal G}$.  For any
pair of arcs $\{a,b\}\subset{\cal A}$, $H_{a,b}=1$ iff $\{a,b\}$ form
a non-backtracking walk of length two, i.e., the head of $a$ coincides
with the tail of $b$, but $b$ is not the reverse of $a$.  

In site percolation on a connected undirected graph $\mathcal{G}$, each 
vertex is chosen to be open with the fixed probability $p$, independent from
other vertices.  We focus on a subgraph
$\mathcal{G}'\subseteq\mathcal{G}$ induced by all open vertices on
$\mathcal{G}$.  For each vertex $v$ on $\mathcal{G}^{\prime}$, let
$\mathcal{C}(v)\subseteq\mathcal{G}'$ be the connected component of
$\mathcal{G}'$ which contains the vertex $v$, otherwise
$\mathcal{C}(v)=\emptyset$.  Denote\cite{Hofstad-2010} by
\begin{equation}
\theta_v\equiv
\theta_v(\mathcal{G},p)=\Pp(|\mathcal{C}(v)|=\infty),
\label{eq:theta-v}
\end{equation}
the probability that $\mathcal{C}(v)$ is infinite.  If $\mathcal{C}(v)$ is infinite, for some
$v$, we say that percolation occurs.  The percolation
transition occurs at the critical probability
$p_c=\sup_p\{p:\theta_v=0\}$.  Similarly, introduce the local
susceptibility,
\begin{equation}
\chi_v\equiv
\chi_v(\mathcal{G},p)=\Ep(|\mathcal{C}(v)|), 
\label{eq:chi-v}
\end{equation}
the expected cluster size connected to $v$, and the associated
critical value $p_T=\inf\{p:\chi_v=\infty\}$.  Generally,
$p_c\le p_T$; on quasitransitive
graphs the two thresholds coincide\cite{Menshikov-1986}.  A third critical value,
$p_u$, corresponds to a transition associated with the number
of infinite clusters. For $p>p_u$ there can be only one infinite
cluster and in general $p_u\ge p_c$.  This inequality is strict on
non-amenable graphs\cite{Haggstrom-Jonasson-2006}.  The uniqueness
phase can be characterized by the connectivity,
\begin{equation}
  \label{eq:connectivity}
  \tau_{u,v}\equiv
  \tau_{u,v}(\mathcal{G},p)=\Pp\bigl(u\in
  \mathcal{C}(v)\bigr),   
\end{equation}
the probability that vertices $u$ and $v$ are in the same cluster.
Indeed, if the percolating cluster is unique, for $p>p_u$, the
connectivity is bounded from below, $\tau_{u,v}\ge \theta_u \theta_v$.

For any non-negative matrix $H$ (finite or infinite) we define 
$p$-norm growth, 
\begin{equation}
  \label{eq:bargr}
  \gr_p H\equiv \sup_v\Bigl\{\lambda>0:
    \liminf_{m\to\infty} {\|e_v^TH^m\|_p\over \lambda^m} =0\Bigr\},
\end{equation}
and a similarly defined $\grsup_p H$ using limit superior.  Here $e_v$
is a vector with the only non-zero element at $v$ equal to one. We
note that for any finite graph, $\gr_p H=\grsup_p H=\rho(H)$.
Moreover, if $H$ is the Hashimoto matrix associated with a tree
${\cal T}$, $\|H^m e_v\|_1$ is the number of sites reachable in $m$
non-backtracking steps from the arc $v\in{\cal A}({\cal T})$.  Then,
$\gr_1 H$ is exactly the growth of the tree\cite{Lyons-1990}, and
$\grsup_1 H$ is the uniformly limited
growth\cite{Angel-Friedman-Hoory-2015}.  Furthermore, on a tree,
$\gr_2 H=(\gr_1 H)^{1/2}$ is the point spectral
radius\cite{Lyons-1990}.  More generally, for any graph ${\cal
  G}$, $\grsup_2 H$ gives an upper bound for the spectral radius
$\rho_{l^2}(H)$ of
$H$ treated as an operator on $l^2({\cal
  A})$; it satisfies the following inequalities
\begin{equation}
(\gr_1 H)^{1/2}\le \rho_{l^2}(H)\le \grsup_2 H\le \grsup_1
H,\label{eq:gr2-bounds}
\end{equation}
where the rightmost inequality is strict if ${\cal G}$ is non-amenable.

\sect{Results.} We prove the following bounds:
\begin{theorem}
  Consider site percolation on a locally finite graph $G$
  characterized by the Hashimoto matrix $H$. Then
  $p_T\ge 1/\grsup_1H$, $p_c\ge 1/\gr_1H$.
\end{theorem}
The first inequality is obtained by evaluating a union bound for $\chi_v$
over all non-backtracking walks starting with $v$
\cite{Hamilton-Pryadko-SIAM-2015,Hamilton-Pryadko-arXiv-2015}; the
second by using the bound on the percolation transition on a graph in
terms of the transition on the universal
cover\cite{Hamilton-Pryadko-PRL-2014}.  The following connectivity
bound follows directly from the alternative definition of
$\rho_{l^2}(H)=\lim_{m\to\infty}\|H^m\|_2^{1/m}$:
\begin{theorem}
  \label{th:connectivity-bound-infinite}
  Consider site percolation on an infinite graph $\mathcal{G}$ with
  maximum degree $d_\mathrm{max}$, characterized by the Hashimoto
  matrix $H$ with $\rho\equiv \rho_{l^2}(H)$.  Then, if $p<1/\rho$,
  connectivity between any pair of sites decays exponentially with the
  distance, i.e., there exists a base $\rho'<1$ and a constant
  $C\ge d_\mathrm{max}(1-p\rho)^{-1}$ such that%
  \begin{equation}
    \label{eq:tau-bound-general}
\forall {\{u,v\}\subset{\cal V}({\cal G})},\quad     \tau_{u,v}\le C
(\rho')^{d(u,v)}.  
  \end{equation}
\end{theorem}

We say that an OLG of a connected graph ${\cal G}$ is strongly
$\ell$-connected, if for any arc $a\in{\cal A}({\cal G})$, there is a
non-backtracking walk of length at most $\ell$ from $a$ to its reverse,
$\bar a$.  When such a graph is finite, the
ratios of the components of the Perron-Frobenius vector of $H$
corresponding to any pair of mutually reverted arcs are uniformly
bounded (up to a constant).  This gives  
\begin{theorem}
  \label{th:subgraph-sequence-local}
  Consider site percolation on a finite graph ${\cal G}$ whose OLG is
  locally strongly $\ell$-connected.  Let $H$ be the Hashimoto matrix
  of ${\cal G}$.  Then, if $\lambda\equiv p\rho(H)<1$, the connectivity
  between any pair of vertices satisfies
  \begin{equation}
    \label{eq:connectivity-bound-finite}
    \tau_{i,j}\le \max(\deg i, \deg j){1+[\rho(H)]^{\ell}\over
      1-\lambda}\,\lambda^{d(i,j)}. 
  \end{equation}
\end{theorem}

Moreover, for any locally-finite graph ${\cal G}$ whose OLG is
locally strongly $\ell$-connected, we have:

\begin{theorem}
  Consider an increasing sequence of subgraphs
  ${\cal G}_t\subset{\cal G}_{t+1}\subset {\cal G}$ convergent to a
  locally-finite graph ${\cal G}$.   The following limit exists
  \begin{equation}
    \label{eq:rho-limit}
    \rho_0\equiv \lim_{t\to\infty}\rho(H_t)\le \rho_{l^2}(H). 
  \end{equation}
  The upper bound is saturated, $\rho_0=\rho_{l^2}(H)$, 
  if the OLG of ${\cal G}$ is locally strongly
  $\ell$-connected.
\end{theorem}

The same parameter $\rho_0$ also defines a lower bound on the uniqueness
transition:
\begin{theorem}
  \label{th:nice-uniqueness-bound} 
  For a locally finite graph ${\cal G}$, the uniqueness transition
  satisfies $p_u\ge 1/\rho_0$.
\end{theorem}
This follows from a bound on the expected number of self-avoiding
cycles passing through a given arc, and the related analysis of
cluster stability\cite{Hamilton-Pryadko-arXiv-2015}.

\begin{example}
  \label{ex:convergence-to-tree}
  A degree-$d$ infinite tree $\mathcal{T}_d$ can be obtained as a
  limit of an increasing sequence of its subgraphs,
  $t$-generation trees ${\cal G}_t=\mathcal{T}_d^{(t)}$.  We have
  $\rho(H_t)=0$ for any $t$, thus $\rho_0=0$, consistent with the
  known fact that there is no uniqueness phase for percolation on
  ${\cal T}_d$.
\end{example}

\sect{Conclusions.} We give lower bounds for all three
transitions usually associated with site percolation on infinite
graphs.  We also identify a region of $p$ where connectivity decays
exponentially with the distance.  For certain graphs with many short
cycles, we give an improved upper bound on connectivity's exponential
falloff with the distance, with explicitly specified parameters.

\sect{Acknowledgments.}
We are grateful to  N.~Delfosse for enlightening
discussions.  This work was supported in
part by the U.S.\ Army Research Office under Grant No.\
W911NF-14-1-0272 and by the NSF under Grant No.\ PHY-1416578.  LPP
also acknowledges hospitality by the Institute for Quantum Information
and Matter, an NSF Physics Frontiers Center with support of the Gordon
and Betty Moore Foundation.

\bibliographystyle{abbrv}
\bibliography{ns16,percol}

\end{document}